\documentclass[natbib]{svjour3}    
\journalname{Celestial Mechanics and Dynamical Astronomy}
\usepackage{amsmath,amsfonts,amssymb,amscd,xcolor,wasysym,mathrsfs}
\usepackage[english]{babel}
\usepackage{natbib} 
\usepackage{url}
\usepackage{mathtools} 
\usepackage{gensymb}
\usepackage{bm}

\newcommand{\GG}[1]{}

\newcommand{\bv}{\boldsymbol}

\def\Z{\mathbb{Z}}

\def\H{\mathcal H}
\def\K{\mathcal K}

\let\e=\varepsilon

\newcommand{\be}{\begin{equation}}    
\newcommand{\ee}{\end{equation}}
\newcommand{\ba}{\begin{eqnarray}}
\newcommand{\ea}{\end{eqnarray}}
\newcommand\beq[1]{ \begin{equation}\label{#1} }
\newcommand{\eeq}{ \end{equation} }

\newcommand\beqa[1]{ \begin{eqnarray} \label{#1}}
\newcommand{\eeqa}{ \end{eqnarray} }
\newcommand{\beqano}{ \begin{eqnarray*} }
\newcommand{\eeqano}{ \end{eqnarray*} }

\newcommand{\nn}{\nonumber \\}

\begin{document}
\title{Dynamical Stability of the Laplace Resonance}

\author{Giuseppe Pucacco\\
}


\institute{              G. Pucacco \at Dipartimento di Fisica and INFN -- Sezione di Roma II,
Universit\`a di Roma ``Tor Vergata", \\
Via della Ricerca Scientifica, 1 - 00133 Roma\\
\email{pucacco@roma2.infn.it} }
\maketitle

\begin{abstract}
We analyse the stability of the \sl de Sitter equilibria \rm in multi-resonant planetary systems. 
The de Sitter equilibrium is the dynamical state of the Laplace resonance in which all  
resonant arguments are librating. The sequence of equilibria exists all along 
the possible states balancing resonance offsets and forced eccentricities.
Possible additional new-de Sitter 
equilibria may exist when at least one of the forced eccentricities is large (the paradigmatic case is Gliese-876). 
In the present work, these families of equilibria are traced up to crossing exact commensurability, where approximate first-order 
solutions diverge. Explicit exact location of the equilibria are determined allowing us to verify the Lyapunov stability of the standard de Sitter equilibrium and 
of the stable branches of the additional ones.


\end{abstract}

\section{Motivations}
\label{intro}

In the catalog of exo-planetary systems, there is a small but constantly growing set of \sl multi-resonant \rm items, 
namely planetary systems with three or more commensurate periods. For example, for 1+3--body systems, there is 
a certain number of them with period ratios given by $n_2/n_1=(k_1-1)/k_1$ and $n_3/n_2$ $ = (k_2-1)/k_2, \; k_1,k_2 \in \Z$ \citep{PiBaMo}, 
configurations that we can call \sl Laplace-like resonances \rm as generalisations of the classical Galilean-system dynamics \citep{CePaPuChao}. The standard Laplace resonance is produced by $k_1=k_2=2.$ 

Since the formation scenarios and subsequent migration of just formed planets 
favour the birth of resonant chains, 
we would expect that the fraction of systems trapped in such states could be quite high, 
especially for systems in which {\it Super-Earths} or {\it mini-Neptunes} are abundant \citep{PePiDaJo}. 
However this is not the case, since the examples of these multi-resonant systems amount to some tens 
over a total of a few thousand systems. Both for transiting planets and for radial-velocity ones, 
the period ratios are known with high precision, therefore the lack of such configurations 
seems to be not an arti-fact \citep{CTL23,NGB24}. A simple explanation of this phenomenon could be that these systems, after 
the disappearance of the proto-planetary disk, are quite prone to dynamical 
instabilities making them short lived systems \citep{Isi17}. 
However, as a matter of fact, the analysis of the stability of Laplace-like resonant configurations 
still present some problematic issues. In fact, numerical simulations appear to show 
that these states are indeed dynamically stable for initial conditions quite close to resonant equilibria and therefore 
other instability mechanisms such as higher-order resonances have been invoked \citep{PiMo,GBM22}. 

However, no analytical proof of dynamical stability is available and, even more, model predictions are at odds with  numerical simulations. In particular, a paradox concerning stability has to be solved: classical works \citep{YoPe,HeCons} predict {\sl instability} beyond a certain threshold of proximity to the resonance. On the other hand, direct numerical integrations point out dynamical stability. Pichierri and Morbidelli (2020) claim that {\it ``the purely resonant system ... with initial condition at vanishing amplitude around a resonant equilibrium point is (Lyapunov) stable for all planetary masses''}. Therefore, it seems that a thorough analysis of the stability nature of the standard Laplace resonance is 
worthy of being endeavored, also in view of the renovated interest in the Galilean system \citep{FeBook2021,La-C,juicemission,lari2024}. 

In the present work we see how to reconcile these predictions by exploiting an accurate location of the `de Sitter-Sinclair' equilibria. This result is obtained with a more effective way of characterising the proximity to the resonance. In fact, we will follow the sequence of equilibria all along 
the possible states in which resonance offsets and forced eccentricities balance, whereas the usual reckoning of Laplace equilibria proceeds by assuming small forced eccentricities \citep{Si75,He-CM-84,BH}. However, if the resonance offsets are small, this assumption leads to an inconsistency and equilibrium solutions may well be characterised by moderate or even large eccentricities. The determination of these equilibrium configurations has to be extended to a more generic setting: this radically modifies the nature of the equilibria and also leads to the bifurcation of `new-de Sitter' families of equilibria \citep{GP-CMDA-2021}. In this case, under certain conditions, a saddle-node bifurcation generates a stable/unstable pair of additional critical points. The stability analysis will therefore be extended also to these equilibria, of which a paradigmatic example is the GJ-876 exo-planetary system. 

The plan of the paper is as follows: in Sect.\ref{sec:model} we remind the basic Hamiltonian of the multi-resonant self-gravitating system; in Sect.\ref{sec:eql} we present the new solutions for the equilibria and analyse the conditions for their linear stability and bifurcations; in Sect.\ref{sec:nfc} we discuss normalisation around these equilibria;  in Sect.\ref{AppSec} we apply these results to the examples of the Galilean system and of GJ-876;  in Sect.\ref{CSec} we conclude with some general remarks.



\section{The dynamical model}
\label{sec:model}

Here we want to describe the main properties of the Laplace-resonant dynamics. When considered as a \sl relative equilibrium \rm of a reduced Hamiltonian system, the de Sitter equilibrium is the dynamical state of the Laplace resonance in which all  
resonant arguments are librating. In real systems (e.g. the standard case characterising the Galilean system), since one of the free eccentricities is larger than the forced one, usually the corresponding argument rotates. However, this state shares the same dynamical properties as the de Sitter one.

One of the easiest way to characterise the resonant 
regime is to exploit a simple model of the libration of the (generalised) Laplace angle. Therefore, the tool we should construct 
is a normal form giving the 
phase plot of a (possibly integrable) reduced dynamics. Preliminary steps for this are the introduction of a suitable coordinate set and the reduction of the Hamiltonian model. Referring to the set of canonical coordinates useful for this purpose, 
we perform some modification with respect to the standard approach adopted  
when describing the Galilean system \citep{He-CM-84,malhotra-icarus-1991}. 

The basis of the work is a simplified model aimed at capturing the essential features of the true system. 
We consider a planar system in which 3 relatively small bodies with orbital elements $a_j, e_j, \lambda_j, p_j, j=1,2,3,$ revolve around a `large' central one. The model Hamiltonian is then given by the Keplerian part plus the resonant terms describing the coupling of the `satellites'
\begin{equation}\label{HKR}
H (L_j, P_j, \lambda_j, p_j) = H_{kep} + H_{res}, \quad (j=1,2,3),
\end{equation}
which survive after averaging with respect to fast angle-combinations. Therefore, the modified Delaunay variables $L_j, P_j, \lambda_j, p_j$ can be interpreted as the osculating elements of the starting model Hamiltonian \citep{CMBR18}. In the present section we illustrate the structure of Hamiltonian \eqref{HKR} and the canonical transformations useful to unveil its dynamics.

 \subsection{Keplerian part}

The Keplerian part of the Hamiltonian is given by:
\begin{equation}
H_{kep} = -\frac{1}{2} \sum_{j=1}^{3} \frac{{M_j}^2 {\mu_j}^3}{{L_j}^2}
\end{equation}
where we define
\begin{equation}
\begin{aligned}
&M_1 = m_0 + m_1, \qquad \mu_1 = \frac{m_0 m_1}{M_1}, \\
&M_2 = M_1 + m_2, \qquad \mu_2 = \frac{M_1 m_2}{M_2}, \\
&M_3 = M_2 + m_3, \qquad \mu_3 = \frac{M_2 m_3}{M_3}.
\end{aligned}
\end{equation}
Units are such that Newton gravitation constant is set to unity. By expanding $H_{kep}$ up to second order with respect to the nominal resonant values $\overline{L}_j$ we get:
\begin{equation}
H_{kep} = \sum_{j=1}^{3} \bigg( \overline{n}_j (L_j - \overline{L}_j) - \frac{3}{2} \eta_j (L_j - \overline{L}_j)^2 \bigg),
\end{equation}
where
\begin{equation}
\overline{n}_j = \sqrt{\frac{M_j}{{\overline{a}_j}^3}} = \frac{{M_j}^2 {\mu_j}^3}{{\overline{L}_j}^3},  \quad
\eta_j = \frac{\overline{n}_j}{\overline{L}_j}.
\end{equation}
We use exact-commensurability conditions in order to compute nominal values, following \cite{bat-mor-AJ-2013} and \cite{GP-CMDA-2021}, so that $\overline{a}_j$ are 
the normalised resonant semi-axes values, $\overline{n}_j$ the nominal mean motions and, in the general case of first-order resonances,
\begin{equation}
\begin{aligned}
k_1 \overline{n}_2 &= (k_1 - 1) \overline{n}_1, \quad k_1=2,3,...\\
k_2 \overline{n}_3 &= (k_2 - 1) \overline{n}_2, \quad k_2=2,3,.... \label{k1k2}\\
\end{aligned}
\end{equation}
Using these relations we can compute the nominal resonant values, in particular this is the choice for the nominal values $\overline{L}_j$ and $\eta_j$. 
 


\subsection{The resonant coupling terms}

The disturbing function, as usual in the case of 
first-order resonances \citep{FeBook2007,Pa15,PiBaMo}, is limited to low-order terms in the 
expansion in the eccentricities.  After averaging with respect to non-resonant `fast' angles, we obtain terms of the form
\begin{equation}
\label{eq:pertHam1}
\begin{aligned}
H_{res}^{(1)}&=-\frac{m_1m_2}{a_2}
\big(f_{12,1} e_1\cos(k_1 \lambda_2- (k_1 - 1) \lambda_1+p_1) \\
&+
f_{12,2} e_2\cos(k_1\lambda_2- (k_1 - 1) \lambda_1+p_2)\big)\ \\
&-{{m_2m_3}\over a_3}\big(f_{23,1} e_2\cos(k_2\lambda_3- (k_2 - 1) \lambda_2+p_2) \\
&+f_{23,2} e_3\cos(k_2\lambda_3- (k_2 - 1) \lambda_2+p_3)\big)\, 
\end{aligned}
\end{equation}
for the first-order terms and 
\begin{equation}
\label{eq:pertHam2}
\begin{aligned}
H_{res}^{(2)} &= -\frac{m_1m_2}{a_2}\big(f_{12,6} e_1 e_2 \cos(p_1-p_2) + f_{12,5} e_1 e_2 \cos(2 k_1 \lambda_2- 2(k_1 - 1) \lambda_1 + p_1 + p_2) \\
&+ f_{12,3} e_1^2 \cos(2 k_1 \lambda_2- 2(k_1 - 1) \lambda_1 + 2p_1) + f_{12,4} e_2^2 \cos(2 k_1 \lambda_2- 2(k_1 - 1) \lambda_1 + 2 p_2)\big) \\[.1cm]
&-\frac{m_2m_3}{a_3}\big(f_{23,6} e_2 e_3\cos(p_2 - p_3) + f_{23,5} e_2 e_3 \cos( 2 k_2\lambda_3 -2(k_2 - 1) \lambda_2 + p_2 + p_3) \\
&+ f_{23,3} e_2^2 \cos(2 k_2\lambda_3 -2(k_2 - 1) \lambda_2 + 2p_2) + f_{23,4} e_3^2 \cos(2 k_2\lambda_3 -2(k_2 - 1) \lambda_2 + 2p_3)\big) \\[.1cm]
& -\frac{m_1m_3}{a_3}\big(f_{13,4} e_1 e_3\cos(p_1 - p_3) \big)\,,
\end{aligned}
\end{equation}
for the 2nd-order terms. The Laplace coefficients $f_{ij,\nu}, i,j=1,2,3, \nu=1,...,6,$  are quantities of order one weakly dependent on the semi-major axis ratios \citep{mur-der-2000}. 

\subsection{Variables adapted to the resonance}
Let us consider the resonance defined by \eqref{k1k2}.
In order to implement the location of relative equilibria and the evaluation of their stability, we introduce a new set of canonical variables {\it adapted} to the first-order resonance. We closely follow the derivation in \cite{GP-CMDA-2021} generalising for completeness to other first-order resonances.

The new set of coordinates is provided by the 
transformation 
$$(L_j, P_j, \lambda_j, p_j), j=1,2,3 \longrightarrow (Q_{\nu}, q_{\nu}), \nu =1,\dots,6.$$ 
The new actions $Q_{\nu}$ are given by:
\beqa{NVeq}
Q_1 &=& P_1\nn Q_2 &=& P_2 \nn Q_3 &=& P_3 \nn
Q_4 &=& \frac{k_1}{3 (k_1 - 1)} L_1 + \frac{L_2}{3} + \frac{L_3}{3 k_2}\left(k_2 - 4 \right) \label{NV1} \\
Q_5 &=& \frac{k_1 + k_2 - 1}{3 (k_1 - 1)} L_1 + \frac{L_2}{3} + \frac{1}{3} (k_2 - 1) (P_1 + P_2 + P_3) \nn
Q_6 &=& L_1 + L_2 + L_3 - P_1 - P_2 - P_3. \nonumber
\eeqa
and the new angles $q_{\nu}$ are 
\begin{equation}
\begin{aligned}
q_1 =& k_1 \lambda_2 - (k_1 - 1) \lambda_1 + p_1 \\
q_2 =& k_1 \lambda_2 - (k_1 - 1) \lambda_1 + p_2 \\
q_3 =& k_1 \lambda_2 - (k_1 - 1) \lambda_1 + p_3 \\
q_4 =& (k_1 \lambda_2 - (k_1 - 1) \lambda_1) - (k_2 \lambda_3 - (k_2 - 1) \lambda_2) \label{NV6} \\
=& (1 - k_1) \lambda_1 + (k_1 + k_2 - 1) \lambda_2 - k_2 \lambda_3 \\
q_5 =& \frac{4 (k_1 - 1)}{k_2} \lambda_1 - \frac{4 (k_1 - 1) + k_2}{k_2} \lambda_2 + \lambda_3 \\
q_6 =& \frac{(k_1 - 1) (k_2 - 4)}{3k_2} \lambda_1 \\
         & - \frac{(k_2 - 4) (k_1 + k_2 - 1)}{3k_2} \lambda_2 + \frac{1}{3} (k_2 - 1) \lambda_3.
\end{aligned}
\end{equation}
In the model at hand, angles $q_5,q_6$ are \sl cyclic \rm (they do not appear anymore in the Hamiltonian \eqref{HKR}) 
and this implies that $Q_5$ and $Q_6$ are exact integrals of motion of $H$.
$$Q_6 =  \sum_{j=1}^{3} (L_j - P_j)$$ 
is the {\it total angular momentum} of the system. We can also use the linear combination
$$ Q_L = 3 Q_5 + (k_2 - 1) Q_6 = \frac{k_1 k_2}{k_1 - 1} L_1 + k_2 L_2 + (k_2 - 1) L_3 $$
which is a {\it scaling} invariant (\cite{bat-mor-AA-2013}, also dubbed {\it spacing parameter} by \cite{MBFM08}). Introducing the {\it angular momentum deficit} \citep{LaskarP}
\beq{amddef}
\Gamma= \sum_{j=1}^{3} P_j = \sum_{j=1}^{3} Q_j,
\eeq
the inverse transformation for the $L_j$ actions is given by:
\begin{equation}
\begin{aligned}\label{IT1}
L_1 = & -(k_1 - 1) \Gamma - (k_1 - 1) Q_4 \\
          & + \frac{4 (k_1 - 1)}{k_2} Q_5 + \frac{(k_1 - 1) (k_2 - 4)}{3k_2} Q_6 \\
L_2 = & k_1 \Gamma + (k_1 + k_2 - 1) Q_4 \\
          & - \frac{4 k_1 + k_2  - 4}{k_2} Q_5 - \frac{(k_2 - 4)(k_1 + k_2 - 1)}{3k_2} Q_6 \\
L_3 = & - k_2 Q_4 + Q_5 + \frac{1}{3} (k_2 - 1) Q_6 . 
\end{aligned}
\end{equation}

The set of variables adapted to the resonance given by \eqref{NV1}-\eqref{NV6}, when compared 
with the original Henrard-Malhotra one \citep{He-CM-84,malhotra-icarus-1991}, has the advantage that 
the action $Q_4$, conjugated to the Laplace argument $q_4$, does not depend on the eccentricities. This fact allows us to decouple the dynamical description of the Laplace libration from those of the free eccentricities around the forced ones, as will be clear with the use of the normal form. We remark that the transformation \eqref{NV1}-\eqref{NV6} is equivalent to that introduced by \cite{Deli17}. 
Still, the main difference relies on the choice of $Q_4$ conjugated 
with the Laplace argument. The conserved quantities in the two approaches 
are connected by linear relations.

The model Hamiltonian can therefore be expressed as
\beq{HLQ}
H (Q_a,q_a; Q_5,Q_6)= \sum_{n=0}^{\infty} H_n  (Q_a,q_a), \quad  a=1,...,4 ,
\eeq
where 
the first three terms in the Hamiltonian \eqref{HLQ} are  
\beqa{HFSM}
H_0&=& \kappa_1 \Gamma + \kappa_4 Q_4 \ ,\label{HMzeroM}\\
H_1&=& -\frac32 A \Gamma^2 - 3 B \Gamma Q_4  -\frac32 C Q_4^2,\label{HMunoM}\\
H_2&=& - \alpha \sqrt{2 Q_1} \cos q_1 - \beta_1 \sqrt{2 Q_2} \cos q_2 \nn
       && - \beta_2 \sqrt{2 Q_2} \cos (q_2 - q_4) - \gamma \sqrt{2 Q_3} \cos (q_3 - q_4),\label{HMdueM}
\eeqa
where
$$\Gamma = Q_1+Q_2+Q_3.$$
Terms of higher order in the expansion $H_n, n>2$ can be added, but we will 
only use them in some later applications. 
The frequencies $\kappa_1(Q_5, Q_6),\kappa_4(Q_5, Q_6),$ are
\beqa{FRM}
\kappa_1 &=& \kappa_{11} Q_5 + \kappa_{12} Q_6 , \label{MSFR1} \\
\kappa_4 &=& \kappa_{41} Q_5 + \kappa_{42} Q_6, \label{MSFR2}
\eeqa
where 
\beqa{NMFRM}
\kappa_{11} &=& 12 \frac{(k_1 - 1)^2}{k_2} \eta_1 + \frac{3 k_1 (k_2 + 4 k_1 - 4)}{k_2} \eta_2 \\
\kappa_{12} &=& \frac{(k_2 - 4) (k_1 - 1)^2}{k_2} \eta_1 + \frac{k_1 (k_2 - 4) (k_1 + k_2 - 1)}{k_2} \eta_2 \\
\kappa_{41} &=& \frac{12 (k_1 - 1)^2}{k_2} \eta_1 + \nn 
&& \frac{3 (4 k_1+ k_2 -4)(k_1 + k_2 - 1)}{k_2} \eta_2 + 3 k_2 \eta_3 \\
\kappa_{42} &=& \frac{(k_1 - 1)^2 (k_2 - 4)}{k_2} \eta_1 + \nn
&& \frac{(k_1 + k_2 - 1)^2 (k_2 - 4)}{k_2} \eta_2 +k_2 (k_2 - 1) \eta_3.
\eeqa
The Keplerian expansion parameters $A,B,C$ are given by
\begin{equation}
\begin{aligned}
A =& (k_1 - 1)^2 \eta_1 + {k_1}^2 \eta_2 \\
B =& (k_1 - 1)^2 \eta_1 + k_1 (k_1 + k_2 - 1) \eta_2 \\
C =& (k_1 - 1)^2 \eta_1 + (k_1 + k_2 - 1)^2 \eta_2 + {k_2}^2 \eta_3
\end{aligned}
\end{equation}
and the coupling parameters $
\alpha , \beta_1 , \beta_2 , 
\gamma $ are defined as usual 
\begin{equation}
\begin{aligned}
&\alpha = {\overline{m}}_2 \epsilon_2 \frac{f_{12,1}}{{\overline L_2}^2 \sqrt{\overline L_1}}\\
&\beta_1 = {\overline{m}}_2 \epsilon_2 \frac{f_{12,2}}{{\overline L_2}^2 \sqrt{\overline L_2}} \\
&\beta_2 =  {\overline{m}}_2 {{\overline{m}}_3}^2 \epsilon_3 \frac{f_{23,1}}{{\overline L_3}^2 \sqrt{\overline L_2}}\\
&\gamma = {\overline{m}}_2 {{\overline{m}}_3}^2 \epsilon_3 \frac{f_{23,2}}{{\overline L_3}^2 \sqrt{\overline L_3}}
\end{aligned}
\end{equation}
with the specific value of the Laplace coefficients and the definitions
\begin{equation}
\begin{aligned}
&\epsilon_k = \frac{m_k}{m_0}, \qquad k= 1,2,3 \\
&{\overline{m}}_j = \frac{m_j}{m_1} = \frac{\epsilon_j}{\epsilon_1}, \qquad j=2,3.
\end{aligned}
\end{equation}
This set has to be considered as a `fixed' set of constant parameters determined by physical quantities (e.g. the masses) 
and the nominal resonant variables. 
Rather, $Q_5$ and $Q_6$ depend on {\it initial conditions} $L_j^0,Q_j^0, \; j=1,2,3,$ which clearly are close but {\it not} coinciding 
with the nominal ones $L_j = \overline L_j, Q_j=0$. 
However, the osculating elements used to compute the integrals of motion $Q_5,Q_6$ may well be substantially displaced with respect to those corresponding to exact commensurability. Choosing for example almost circular orbits would imply semi-major axis ratios sensibly different from the nominal ones as suggested by typical formation scenario driven by dissipative interactions \citep{GBM22,NGB24}. 

From the analytical point of view, \eqref{HLQ} is a \sl reduced Hamiltonian \rm \citep{AbMa}. Its dynamics occur on a 4-DoF (Degrees of Freedom) reduced phase-space \citep{He-CM-84,BH}. In particular, its critical points provide equilibria which, when mapped back through the inverse transformation \eqref{IT1}, in combination with \eqref{NVeq}, give periodic orbits. These orbits inherit the stability nature of the parent equilibrium. In the next section we will introduce a unique parameter of `proximity' to the resonance which allows us to properly parametrise the status of the reduced systems. We remark that the ordering of the terms in \eqref{HLQ} reflects only their meaning and origin. The actual ordering in the Hamiltonian expansions and the normal form construction will be established case by case on the basis of suitable assumptions. 

\section{Equilibria of the Laplace-resonance}
\label{sec:eql}

Historically, the description of the Laplace-resonant state as an equilibrium can be attributed to \cite{desitter-1909,desitter-MNRAS-1931}. The approach was later extended by \cite{Si75} and discussed in several other works. New `de Sitter-Sinclair' equilibria may appear 
with different apsidal combinations and possibly higher forced eccentricities \citep{GP-CMDA-2021,WZB}.

We have a reduced 4-DoF Hamiltonian vector field associated to \eqref{HLQ}. Fixed points of this field give the equilibria of the system. In the first-order (in eccentricity) case, an explicit 
expression of these critical points is straightforward \citep{Si75,GP-CMDA-2021}. It would be nice to have analytical solutions in the general case,
since they provide clues for a full understanding of what is going on. 
We remark that from now on, we limit all expressions to the `standard' Laplace-resonance with 
$$k_1 = k_2 = 2 $$
so that, as usual
\ba\label{cootr}
q_1&=&2\lambda_2-\lambda_1+p_1 \ , \quad\quad Q_1 = P_1 \ ,\label{q1}\\
q_2&=&2\lambda_2-\lambda_1+p_2 \ , \quad\quad Q_2 = P_2 \ , \label{q2}\\
q_3&=&2\lambda_2-\lambda_1+p_3 \ , \quad\quad Q_3 = P_3 \ , \label{q3}\\
q_4&=&3\lambda_2-2\lambda_3-\lambda_1 \ , \quad \;\;
Q_4 = {\scriptstyle{\frac13}}\left(2 L_1 + L_2- L_3\right) \ , \label{q4}\\
q_5&=&\lambda_1-\lambda_3 \ , \quad\quad\quad\quad \;\;
Q_5 = {\scriptstyle{\frac13}}\left(3L_1+L_2+\Gamma\right) \ , \label{q5}\\
q_6&=&\lambda_3\ , \quad\quad\quad\quad\quad\quad \;\;\,
Q_6 = L_1+L_2+L_3-\Gamma.  \label{q6}
\ea
We remark that the third resonant angle is the only one differing from the standard convention in which is used the combination $q_3-q_4=2\lambda_3-\lambda_2+p_3$. Moreover we can define the `scaling parameter', which is the integral of motion
\beq{SP22}Q_L = 4L_1+2L_2+L_3 = 3Q_5+Q_6.\eeq
 
\subsection{Resonance proximity parameter}
In order to highlight the dynamics of the Laplace libration we need to make a further simplification of our starting model. A standard way 
to simplify the structure of the Hamiltonian is that of constructing a `normal form'. We then investigate the equilibria of the models introduced above 
and proceed with normalising around them. We remark that, along the lines pioneered by \citet{He-CM-84}, the librational normal form 
is not limited to small amplitudes. Before approaching this task, we note that the ordering 
implicit in  (\ref{HMzeroM}-\ref{HMdueM}) is not necessarily appropriate. As a matter of fact, by using a suitable book-keeping parameter $\e$, we can rearrange the terms 
of the reduced Hamiltonian denoting it as 
$$
\H_0 + \e \H_1 + \e^2 \H_2 + ...,
$$
with the freedom of ordering terms on the basis of specific assumptions. 

Let us assume that $Q_4$ (by construction) and $A,B,C$ (by definition) are quantities of order 1. For typical configurations characterised by small eccentricities (e.g. in the Galilean system) we have that $\sqrt{Q_j}  \sim e_j \sim 10^{-3}$; the same order of magnitude can be associated to the 
coupling parameters $\alpha,\beta_1,$ etc. On these bases, a reasonable ordering is
\beqa{HBSM}
\H_0&=& \kappa_4 Q_4 - \frac32 C Q_4^2 \ ,\label{HBzeroM}\\
\H_1&=& (\kappa_1 - 3 B Q_4 ) \Gamma + {\H}_{res}^{(1)}, \label{HBunoM}\\
\H_2&=&  - \frac32 A \Gamma^2 , \label{HBdueM}
\eeqa
where ${\H}_{res}^{(1)}$ is the resonant term \eqref{HMdueM} expressed in the new variables. However, 
this choice, which essentially embodies the assumptions performed in previous works \citep{Si75,YoPe,He-CM-84}, is all the less acceptable the further the eccentricity grows. If eccentricities are, say, one order of magnitude larger, a more sensible ordering is 
\beqa{HBSS}
\H_1&=& (\kappa_1 - 3 B Q_4 ) \Gamma - \frac32 A \Gamma^2, \label{HBunoS}\\
\H_2&=&  {\H}_{res}^{(1)}. \label{HBdueS}
\eeqa
This choice will allow us to get more accurate locations of the equilibria, even if at the price of 
a somewhat more complicated algebra. 

Before proceeding to this task, 
let us perform a preliminary shift which permits an effective strategy to describe the resonance. 
We assume that, projecting the libration of the system around the equilibrium on the phase-space planes $(Q_a,q_a),\;  a=1,...,4 $, we get oscillations of the conjugate pair $(Q_4,q_4)$.
We reserve to check this assumption with the normal form construction of Sect.\ref{sec:nfc}. 
Therefore, we assume that $Q_4$ oscillates with small amplitude around an equilibrium value that, at first 
order, can be approximated by exploiting the zero-order term $\H_0$. We get:
$$
\dot q_4 = 0 \to \frac{\partial \H_0}{\partial Q_4} = 0 \longrightarrow Q_4^{(0)} = \frac{\kappa_4}{3C}.
$$
Therefore we can introduce a new `small' variable 
\beq{Lambdadef}
\Lambda = Q_4 - Q_4^{(0)}
\eeq
and assume that $\Lambda$ too is of order $\e$. The first-order 
term \eqref{HBunoS} in the Hamiltonian can then be rewritten as 
\beq{resoH}
\H_1 = \omega \Gamma - \frac32 \left(A \Gamma^2 + 2 B \Lambda \Gamma + C \Lambda^2 \right),\eeq
where 
\beq{omefr}
\omega (Q_5,Q_6) \doteq \kappa_1 - 3 B Q_4^{(0)} = \kappa_1 - \frac{B}{C} \kappa_4,
\eeq
is a frequency  which has a relevant role in the de Sitter dynamics. Since 
it can profitably be used as a reliable way to measure the distance of the system from exact resonance, we call it {\it resonance proximity parameter} adopting the same expression introduced by \cite{B15} in the case of two resonant planets. By using (\ref{MSFR1}-\ref{MSFR2}) we get
\beq{omefrq}
\omega = \frac3C \left[(\eta_1 \eta_2 + 2 \eta_1 \eta_3 + 4 \eta_2 \eta_3) Q_L - (\eta_1 \eta_2 + 4 \eta_1 \eta_3 + 16 \eta_2 \eta_3) Q_6 \right], 
\eeq
but, recalling that $ L_j = n_j / \eta_j$ and 
$$ Q_L = 4 L_1 + 2 L_2 + L_3, \quad Q_6 = L_1 + L_2 + L_3 - \Gamma,$$ 
it is very enlightening to express it in terms 
of direct observables
\beq{omefro}
\omega = \frac1C \left[(C-B)(2n_2 - n_1)  + B (2 n_3 - n_2) + 2CK \Gamma \right], \quad K \doteq \frac3{2C}(AC-B^2)
\eeq
and delve into its properties with some more details. 

As dynamical and astronomical evidence suggest \citep{PiBaMo,CTL23}, both `resonance offsets' $2n_2 - n_1$ and $2n_3 - n_2$ are most often negative quantities. Moreover, in the case of almost circular orbits, the angular momentum deficit $\Gamma$ is quite small. Therefore, reminding the ordering $C > B > 0$ and the condition $K \sim |{\rm O}(1)|$, it is natural to consider the proximity parameter \eqref{omefro} as a negative quantity well distinct from zero. This is precisely the assumption at the basis of the classical works by \cite{Si75} and \cite{YoPe} leading to first-order estimates for the equilibria \citep{GP-CMDA-2021}. 

However, we remark that, as a matter of principle, $\omega$ can be chosen quite arbitrarily in order to produce a whole family of models. In fact, we observe that $\omega (Q_5,Q_6)$ \sl may well vanish and even become positive. \rm Let us demonstrate this statement from a purely mechanical point of view. After the canonical transformation (\ref{q1})-(\ref{q6}), the original 6-DoF system is {\it reduced} to a 4-DoF Hamiltonian system parametrically dependent on $Q_5,Q_6$. We can indeed imagine initial conditions either produced from a particular previous evolution or even concocted at art which provide values compatible with $\omega \simeq 0$. There are at least two possibilities, not mutually excluding: mean motions can be close to exact commensurability with eccentricities remaining small or, on the other hand, both resonance offsets may stay negative and large with at least one of the forced eccentricities much larger than usual. In both these cases, the proximity parameter may vanish and even become positive. 

The sequence of equilibria is then the set of states balancing resonance offsets and forced eccentricities. In the following, we will analyse these cases and show that they are indeed concrete possibilities useful in applications. We stress that, whatever these cases occur, they correspond to some given values of starting $L_j^0,Q_j^0$ and, in turn, of $Q_5,Q_6$ which uniquely specify a reduced system (in addition to physical parameters). 

Actually, the simplification achieved considering a one-parameter family can also be  achieved by normalising action variables by one of the constant of motion (usually the spacing parameter $Q_L$, as done e.g. in \cite{CouRoCo}). Using $\omega$ as a proximity parameter to the resonance has the advantage to maintain a greater generality in the description of the dynamics. 

\subsection{Equilibrium solutions and the bifurcations of the `New-de Sitter' equilibria}
We can now determine the equilibrium sequence keeping, as in \eqref{resoH} also the quadratic terms in the actions. For this, we substitute Poincar\'e coordinates given by:
\begin{equation}
x_i = \sqrt{2 Q_i} \cos{q_i}, \qquad y_i = \sqrt{2 Q_i} \sin{q_i} , \quad i=1,2,3, \label{PVD}
\end{equation}
so that 
$$ \Gamma = \frac12 \sum_i (x_i^2 + y_i^2) $$
and, in order to harmonise the notation, define $\lambda = q_4$.
Therefore, we now work with the resonant Hamiltonian
\beq{resoH2}
\H (\Lambda, x_i, \lambda, y_i)= \omega \Gamma - \frac32 \left(A \Gamma^2 + 2 B \Lambda \Gamma + C \Lambda^2 \right) + 
\H_{res}^{(1)}   \eeq
where $\H_{res}^{(1)}$ is the resonant coupling part \eqref{HMdueM} expressed in the new variables:
\begin{equation}\label{HR1P}
\begin{aligned}
\H_{res}^{(1)} (x_i, y_i, \lambda) = - \alpha x_1 - \beta_1 x_2 - \beta_2 (x_2 \cos{\lambda} + y_2 \sin{\lambda}) - \gamma (x_3 \cos{\lambda} + y_3 \sin{\lambda}).
\end{aligned}
\end{equation}
To consider in \eqref{resoH2} $\H_{res}^{(1)}$ of the same order as $\H_1$ is a bit in contrast with the new book-keeping imposed by  (\ref{HBunoS}-\ref{HBdueS}), but is nonetheless essential to get meaningful equations for the critical points.

Let us denote with $X_i,Y_i,\Lambda_E,\lambda_E$ the equilibrium values of the eccentricity vectors $x_i, y_i$ and of the pair $\Lambda,\lambda$. 
From Hamiltonian \eqref{resoH2}, the equilibrium values are given by solving the systems of equation 
\beqa{Lld}
\dot{\Lambda} &=& 0 \to \frac{\partial \H}{\partial \lambda} = 0 \longrightarrow \beta_2(- x_2 \sin{\lambda} + y_2 \cos{\lambda}) + \gamma (-x_3 \sin{\lambda} + y_3 \cos{\lambda}) = 0 \label{Lambdadot}\\
\dot{\lambda} &=& 0 \to \frac{\partial \H}{\partial \Lambda} = 0 \longrightarrow C \Lambda + B  \Gamma = 0 \label{lambdadot}
\eeqa
and $\dot{x}_i = 0$ and $\dot{y}_i = 0$ respectively giving
\begin{equation}
\begin{aligned}
& (3 A \Gamma + 3 B \Lambda - \omega) y_1  = 0 \\
& (3 A \Gamma + 3 B \Lambda - \omega) y_2 + \beta_2 \sin{\lambda} = 0 \label{XEQ1}\\
& (3 A \Gamma + 3 B \Lambda - \omega) y_3 + \gamma \sin{\lambda} = 0
\end{aligned}\end{equation}
and
\begin{equation}
\begin{aligned}
& (3 A \Gamma + 3 B \Lambda - \omega) x_1 + \alpha  = 0 \\
& (3 A \Gamma + 3 B \Lambda - \omega) x_2 + \beta_1 + \beta_2 \cos{\lambda} = 0 \label{YEQ1}\\
& (3 A \Gamma + 3 B \Lambda - \omega) x_3 + \gamma \cos{\lambda}= 0.
\end{aligned}
\end{equation}
The solutions of the combined set given by \eqref{Lambdadot} and \eqref{XEQ1} are given by $Y_j = 0$ and by $\lambda_{E} = 0,\pi$ so that \eqref{YEQ1} now become
\begin{equation}
\begin{aligned}
& (2K \Gamma_E- \omega) X_1 - \alpha  = 0 \\
& (2K \Gamma_E- \omega) X_2 - \beta_1 \mp \beta_2 = 0 \label{YEQ2}\\
& (2K \Gamma_E- \omega) X_3 \mp \gamma = 0
\end{aligned}
\end{equation}
where $K$ has been defined above in \eqref{omefro}, the {\it equilibrium angular momentum deficit} is 
$$ \Gamma_E = \Gamma\big\vert_{Y_i=0} = \frac12 (X_1^2 + X_2^2 + X_3^2)$$
and it has been exploited the relation 
\be\label{LEGE}
\Lambda_E = -\frac{B}{C} \Gamma_E \ee
solution of the $\dot{\lambda} = 0$ condition \eqref{lambdadot}. Upper and lower signs respectively refer to the `de Sitter' 
($\lambda_{E} = \pi$) and `anti-de Sitter' ($\lambda_{E} = 0$) equilibria.

The exact solutions of the system of equations \eqref{YEQ2} are found in the following way. We assume $X_j \ne 0, \forall j=1,2,3.$ The consistency of this assumption is checked in the end but is justified by the above discussion concerning $\omega$. From the quotient of each equation of the system by $X_j$, we get the chain of equalities 
$$ K (X_1^2 + X_2^2 + X_3^2) = \omega - \frac{\alpha}{X_1}
= \omega - \frac{\beta_1  \mp \beta_2}{X_2}
= \omega - \frac{\mp \gamma}{X_3}. $$
From the last two we get
\be\label{x123}
X_2 = \frac{\beta_1  \mp \beta_2}{\alpha} X_1, \quad
X_3 = \frac{\mp \gamma}{\alpha} X_1,\ee
which, when inserted into the first equality, give
$$
\frac{\alpha}{X_1} = \omega - K (X_1^2 + X_2^2 + X_3^2) = \omega - {\hat K} X_1^2,
$$
where
\be\label{Khat}
{\hat K} = \frac{K(\alpha^2 + (\beta_1  \mp \beta_2)^2 + \gamma^2)}{\alpha^2}.\ee
Therefore, to find exact solutions for the equilibria we have to solve the standard cubic
$$ X_1^3 - \frac{\omega}{\hat K} X_1 + \frac{\alpha}{\hat K} =0. $$
The cubic equation has the three solutions
$$
X_E = \left(-\frac{\alpha}{2\hat K} + \sqrt{\Delta}  \right)^{1/3} + \left(-\frac{\alpha}{2\hat K} - \sqrt{\Delta}  \right)^{1/3}, \quad 
\Delta \doteq \frac{\alpha^2}4 - \frac{\omega^3}{27 {\hat K}}.$$
All of them are real if $\Delta \le 0$. Therefore, interpreting the discriminant as a function of the resonance proximity parameter, the value $\omega_b$ such that $\Delta(\omega_b)=0$ determines a {\it bifurcation} from one to three critical points of the reduced Hamiltonian. Therefore, for
\be\label{omebif}
\omega \ge \omega_b = \left(\frac{27}4 \alpha^2 \hat K  \right)^{1/3} = 
\left(\frac{27K}4(\alpha^2 + (\beta_1  \mp \beta_2)^2 + \gamma^2)  \right)^{1/3},\ee
`New-de Sitter equilibria' will appear \citep{GP-CMDA-2021}. Explicitly, we have
\begin{equation}
\begin{aligned}
X_{1E} =& \left[\frac{\alpha}{2\hat K}\left(-1 + \sqrt{1-{\left(\frac{\omega}{\omega_b} \right)^3}} \right)  \right]^{1/3} + 
                  \left[\frac{\alpha}{2\hat K}\left(-1 - \sqrt{1-{\left(\frac{\omega}{\omega_b} \right)^3}} \right)  \right]^{1/3}, \label{XE1E} \\
X_{2E} =& \frac{\beta_1 \mp \beta_2}{\alpha} X_{1E},\\
X_{3E} =& \pm \frac{\gamma}{\alpha} X_{1E} .
\end{aligned}
\end{equation}
Since $\omega_b > 0$, \eqref{omebif} says that bifurcation occurs only in the quite extreme cases of positive resonance offsets or for large values of forced eccentricities. 

For sake of convenience it can be useful to re-derive the first-order approximation of the relative equilibria of the system. 
Coming back to the original book-keeping providing the ordering of the Hamiltonian as in (\ref{HBzeroM}-\ref{HBdueM}), we get the following conditions for the critical points:
\begin{equation}
\begin{aligned}
& \omega  y_1  = 0\\
& \omega  y_2 - \beta_2 \sin{\lambda}= 0 \label{XEQ11}\\
& \omega  y_3 - \gamma \sin{\lambda} = 0.
\end{aligned}
\end{equation}
and
\begin{equation}
\begin{aligned}
& \omega  x_1 - \alpha  = 0 \\
& \omega  x_2 - \beta_1 - \beta_2 \cos{\lambda} = 0 \label{YEQ11}\\
& \omega  x_3 - \gamma \cos{\lambda} = 0.
\end{aligned}
\end{equation}
The solutions are still
\beq{YE1}
y_i \doteq Y_{i} = 0, \; \forall i=1,2,3,\eeq 
and:
\begin{equation}
\begin{aligned}
x_1 \doteq X_{1}^A =& \frac{\alpha}{\omega} \label{XE1} \\
x_2 \doteq X_{2}^A =& \frac{\beta_1 \mp \beta_2}{\omega} \\
x_3 \doteq X_{3}^A =& \mp \frac{\gamma}{\omega} 
\end{aligned}
\end{equation}
where, as before, the upper sign corresponds to the `de-Sitter' equilibrium value $\lambda_{E} = \pi$ and the lower sign corresponds to the `anti-de-Sitter' equilibrium $\lambda_{E} = 0$. The apex $A$ stands for {\it asymptotic} meaning that, as we can see comparing with exact solutions, they are valid for large values of $|\omega|$.

Since all solutions are characterised by $Y_{i}=0$, from the definitions \eqref{PVD} we see that $q_{iE}=0,\pi$ according to the sign of $X_{i}$, whereas their absolute values provide the forced eccentricities according to
$$
e_i = \frac{|X_{i}|}{\sqrt{\overline L_i}}, \quad i=1,2,3. 
$$
We remark the presence of the frequency \eqref{omefro} in the denominators of the solutions $X_{i}^A$. Therefore, when trying to extend the validity of the solutions to small values of $\omega$ we notice the implicit contradiction of diverging values of the eccentricity breaking the assumption at the basis of the first-order treatment. 
However, for small eccentricities these solutions are quite accurate and essentially coincide with those obtained by \cite{Si75}.

\subsection{Equilibria: an approximate evaluation}
It can be useful to find some simple approximation to the solution \eqref{XE1E}.
We will proceed in two steps of increasing accuracy. 
We check that, if $|\omega| \gg 2 \Gamma_E$, since $K \sim |{\rm O}(1)|$, we recover solutions \eqref{XE1}.
On the other hand, if $\omega = 0$, we get the solutions
\begin{equation}
\begin{aligned}
X_{1}^0 =& -\frac{\alpha}{k^{\mp}} \label{XE10} \\
X_{2}^0 =& -\frac{\beta_1 \mp \beta_2}{k^{\mp}} \\
X_{3}^0 =& \pm \frac{\gamma}{k^{\mp}} ,
\end{aligned}
\end{equation}
where
\be\label{Kaa}
k^{\mp} = \left(K(\alpha^2 + (\beta_1  \mp \beta_2)^2 + \gamma^2)  \right)^{1/3}.
\ee
A simple second-order approximation can be obtained with an exponential fit of these two sets. We get
\be\label{XEXP1}
X_{j} = X_{j}^0 {\rm e}^{\omega/\omega_A}, \quad (j=1,2,3)
\ee
where
\be\label{omegaa}
\omega_A = \frac{k^{\mp}}{\rm e}.
\ee
Solutions \eqref{XEXP1} are a very good approximation of the exact ones for moderate values of $\omega$ around zero.

\subsection{Second-order equilibria: an approximate solution including 2nd-order terms in eccentricity}
We remark that the possibility to get the above closed-form solution is due to truncating the resonant term to first-order in eccentricity. Actually, one (or more) of the forced eccentricities can be so big to violate the book-keeping hierarchy assumed above. In this respect, we are required to 
take into account also 2nd-order terms in the eccentricities, therefore including $\H_{res}^{(2)} (x_i, y_i, \lambda)$,  namely \eqref{eq:pertHam2} expressed in terms of the new coordinates. 
In this more general case, in view of the presence of non-diagonal terms in the system of equations (due to 
the quadratic couplings in the Poincar\'e variables given by \eqref{eq:pertHam2}), it is no longer possible to obtain the explicit solution written above. 
In \cite{GP-CMDA-2021} we looked for solutions of the form $X_1 = X_{1}^{(0)} + \epsilon X_1^{(1)}$ 
by making the hypothesis that the eccentricity of the most internal satellite is much greater than those of the other two (namely $X_1 \gg X_2,X_3$) so that the cubic equation to solve now is
\be
\left(\omega - K X_1^2 \right) X_1 - \epsilon (\alpha + 2 \alpha_2 X_1) =0. \label{C3}
\ee 
To order zero in $\epsilon$, we get the three solutions
$$
 X_1^{(0)} =0
$$
and
$$ X_1^{(0)} = \pm \sqrt{\frac{\omega}{K}}.$$
The first of these provides again $X_1^{(1)} = \alpha/\omega = X_{1}^A$, namely \eqref{XE1} already obtained above. 
But, if the argument of the square root is positive, we obtain two additional sets of solutions:
\ba
X_{1E}^{(N)} &=& \pm \frac{\omega-\alpha_2}{\sqrt{K\omega}} - \frac{\alpha}{2\omega}, \;\; N=2,3. \label{X1N} \nn
X_{2E}^{(N)} &=& \frac{\beta_1  \mp \beta_2 + \alpha_{12} X_{1}^{(N)}}{\omega -2 \beta_{12} - K (X_{1}^{(N)})^2}  \label{X2N} \nn
X_{3E}^{(N)} &=& \frac{\mp \gamma}{\omega -2 \gamma_2 - K (X_{1}^{(N)})^2} \label{X3N}
\ea
Numerically, this approximate set is in substantial agreement with the exact solution \eqref{XE1E} for $|\omega|$ not too small with the advantage of incorporating the 2nd-order coefficients which can play a determining role 
in fixing the signs of the solutions and as a consequence the values ($0$ or $\pi$) of the libration centres. They are:
\ba\label{alfak}
\alpha_2 &\doteq \frac{\overline m_2^2 \e_2 f_{12,3}}{\overline L_1 \overline L_2^2} > 0, \\
\alpha_{12} &\doteq\frac{\overline m_2^2 \e_2}{\sqrt{\overline L_1} \overline L_2^{3/2}} (f_{12,5}+f_{12,6}), \\
\beta_{12} &\doteq
\frac{ \overline m_2^2 \e_2 f_{12,4}}{\overline L_2^{3}} + \frac{\overline m_2 \overline m_3^2 \epsilon_3 f_{23,3}}{\overline L_3^2 \overline L_2}, \\
\gamma_2 &\doteq\frac{\overline m_2 \overline m_3^2 \epsilon_3 f_{23,4}}{\overline L_3^3}.
\ea


\subsection{Linear stability of the de Sitter equilibria}
\label{stabeq}

We can refine the check of the nature of the equilibria 
by evaluating their linear stability. If we obtain linear instability, 
this is sufficient for instability tout-court and we can discard the corresponding solution.

Collectively denoting with $z$ the set $(x_k, y_k,\Lambda,\lambda)$, 
we consider the linear Hamiltonian equations providing the variational system \citep{YoPe}
\be\label{VES}
\frac{d}{dt} \delta z = {  J } H_{zz} \big\vert_E \delta z,\ee
where $J$ is the symplectic matrix. Looking for solutions of the form
$$
\delta z = Z {\rm e}^{\sigma t}, $$
we have to compute the eigenvalues of the Hamiltonian matrix in \eqref{VES}, 
namely the 
solutions of the characteristic equation
$$
\det ( {  J } H_{zz} \big\vert_E - \sigma {  I }) = 0,$$  
where $I$ is the unit matrix. If we find at least a pair of real eigenvalues, the equilibrium is unstable. 

By using the first-order solution \eqref{XE1} we have shown \cite{GP-CMDA-2021} that a pair of eigenvalues 
pass from pure imaginary to real for
\beq{LUNST}
      \omega > \omega_U \doteq - \frac{3B^2}C (X_1^2 + X_2^2 + X_3^2).
     \eeq
     When applied to the Galilean system, the numerical value of $\omega_U$ so obtained is in substantial agreement with the numerical prediction made by \citep{YoPe}. 
     
     However, when implementing any of the exact or approximate forms of second-order solutions, in which there is no pronounced increase of the forced eccentricities around nominal resonance, there is no more trace of change in the nature of the eigenvalues: in the framework of the current model in which the equilibrium solution are provided by \eqref{XE1E}, the  eigenvalues remain pure imaginary confirming the dynamical (Lyapunov) stability of the relative equilibria for any value of the resonance proximity parameter in the range compatible both with formation scenarios and with observations. Due to its algebraic structure, the characteristic equation is no more explicitly solvable so we will check each case with a numerical approach (see Section \ref{AppSec}).
     

\section{Normalisation}
\label{sec:nfc}

The construction of a normal form allows us to get an integrable approximation of the dynamics around the equilibrium. The particular structure of the starting Hamiltonian \eqref{resoH2} requires a transformation which combines averaging and translation: the Lie-transform method \citep{MO02,FeBook2007} provides both with a single operation \citep{He-CM-84}.

Suppose to start with a Hamiltonian
\beqa{oldH}
\mathcal{H} = \mathcal{H}_0 ({\bv Q}) + \e \mathcal{H}_1({\bv Q},{\bv q}) + \e^2 \mathcal{H}_2({\bv Q},{\bv q}) + ...
\eeqa
where $\mathcal{H}_0$ is the \sl unperturbed \rm integrable part and the perturbation 
is given by a series ordered according to suitable assumptions on the coefficients 
of the expansion. 
In the spirit of the Lie-transform method, we pass from old variables $({\bv
Q},{\bv q})$ to new variables $({\bv
Q}',{\bv q}')$ by means of a generating function  $\chi_1({\bv Q}',{\bv q}')$, such that the new Hamiltonian is 
\beqa{newH}
&& \mathcal{H}^{(1)}({\bv Q}',{\bv q}') = \mathcal{H}_0({\bv Q}') +
\varepsilon \left[\mathcal{H}_1({\bv Q}',{\bv
q}')+\{ \mathcal{H}_0({\bv
Q}'),\chi_1({\bv Q}',{\bv q}')\} \right]+\nonumber\\
&&\varepsilon^2 \left[\mathcal{H}_2({\bv Q}',{\bv q}') + \{
\mathcal{H}_1({\bv Q}',{\bv q}'),\chi_1({\bv Q}',{\bv q}')\}+\frac12 \{\{\mathcal{H}_0({\bv
Q}'),\chi_1({\bv Q}',{\bv
q}')\},\chi_1({\bv Q}',{\bv q}')\} \right] \nn
\eeqa
plus higher-order terms. 

We look for a transformation such that the new Hamiltonian is \sl in normal form, \rm namely $\sum_n \e^n\K_n$, 
is constructed so to commute with the integrable part:
\beq{CH0}
{\cal L}_{\mathcal{H}_0} \K_n \doteq \{\K_n, \mathcal{H}_0 \} =0, \quad \forall n.\eeq
At first order in $\e$, \eqref{newH} gives
\begin{equation}\label{eq:HomologicGen1}
\mathcal{H}_1({\bv Q}',{\bv q}')+\{ \mathcal{H}_0({\bv Q}'),
\chi_1({\bv Q}',{\bv q}')\} ={\K}_1({\bv Q}').
\end{equation}
This is solved by equating ${\K}_1$ to those terms of $\mathcal{H}_1$ which are in the kernel of the Hamiltonian operator ${\cal L}_{\mathcal{H}_0}$ and using the remaining terms to integrate the generating function.
Accordingly, at second order in $\e$, we get\footnote{Strictly in \eqref{eq:HomologicGen2} we should distinguish the new variables from $({\bv Q}',{\bv q}')$ denoting them with, e.g., a double prime: however, to not overwhelm the notation, we keep the usual attitude and leave the same symbol for the new variables at each step of normalisation.}
\begin{equation}\label{eq:HomologicGen2}
 \mathcal{H}^{(2)}({\bv Q}',{\bv q}') = \mathcal{H}_2({\bv Q}',{\bv q}') + \frac12 \{
\mathcal{H}_1({\bv Q}',{\bv q}') + {\K}_1({\bv Q}'),\chi_1({\bv Q}',{\bv q}' )\}+
\{\mathcal{H}_0({\bv Q}'),\chi_2({\bv Q}',{\bv q}')\}
\end{equation}
from which ${\K}_2$ and a second generating function $\chi_2({\bv Q}',{\bv q}')$ are computed and so forth at steps $n>2$.

We follow above procedure by constructing a normal form for the Laplace resonance with the two book-keeping schemes discussed above. First we recall the first-order (in the integrable part) one introduced by \cite{GP-CMDA-2021} valid for $\omega$ far from zero. Then we introduce the normalisation scheme for the 2nd-order integrable part able to cope with the case $\omega \sim 0$.

\subsection{Linear normalisation}
Consider again Hamiltonian \eqref{resoH2}. Let us first adopt the book-keeping criterion introduced to treat the small-eccentricity cases (\ref{HBunoM}-\ref{HBdueM}) and so consider the starting Hamiltonian
\beq{resoH1}
\H = \omega\Gamma + \e \H_{res} + \e^2 \left(  - 3 B \Lambda \Gamma - \frac32 C \Lambda^2 - \frac32  A \Gamma^2 \right)  .\eeq
The $\omega$ frequency is associated with the free eccentricity oscillations. We assume it to be `fast' with respect to the libration of the Laplace argument. We can therefore normalise with respect to the `isotropic oscillator' \citep{SVM}
$$
\H_0=\omega \Gamma = \omega (Q_1+Q_2 + Q_3)
$$
by removing the dependence of the Hamiltonian on fast angles. 
From a quick inspection, we see that no term in $\H_1 = \H_{res}$ commutes with $\H_0$. The solution to the first-order homological equation is therefore $\K_1 = 0$ and 
\beqa{oldchi}
\chi_1 ({\bv Q}',{\bv q}') &=& -\frac{\alpha}{\omega} \sqrt{2 Q'_1} \sin{q'_1} 
-\frac{\beta_1}{\omega} \sqrt{2 Q'_2} \sin{q'_2} \nn
&& -\frac{\beta_2}{\omega} \sqrt{2 Q'_2} \sin{(q'_2-\lambda')} -\frac{\gamma}{\omega} \sqrt{2 Q'_3} \sin{(q'_3-\lambda')}. 
\eeqa
At second order, examining \eqref{eq:HomologicGen2}, 
the first Poisson bracket turns out to be equal to
$$
-\frac{\alpha^2 + \beta_1^2 + \beta_2^2 + \gamma^2}{2 \omega} - \frac{\beta_1 \beta_2}{\omega} \cos \lambda',
$$
so that the second homological equation has solutions $\chi_2=0$ and, neglecting a trivial constant term, 
\beq{nfl}
\K_2 = 3 B \Lambda' (Q'_1+Q'_2 + Q'_3) + \frac32 C \Lambda'^2 + \frac32  A (Q'_1+Q'_2 + Q'_3)^2 + \frac{\beta_1 \beta_2}{\omega} \cos \lambda'.
\eeq
The transformed angular momentum deficit can be denoted as
\beq{amddefp}
\Gamma_E=Q'_1+Q'_2 + Q'_3,
\eeq
and, in this approximation, is a conserved quantity. Therefore, the 2nd-order normal form is equivalent to the reduced Laplace Hamiltonian \citep{GP-CMDA-2021}
\beq{rnfl}
K_L (\Lambda', \lambda'; \Gamma_E) = 3 B \Gamma_E \Lambda'  + \frac32 C \Lambda'^2 + \frac{\beta_1 \beta_2}{\omega} \cos \lambda'.
\eeq
It provides a first-order approximation of the libration frequency around the reference $\lambda'=\pi$ equilibrium
\beq{fola}
\omega_L=\sqrt{\frac{3C\beta_1 \beta_2}{\omega}}\eeq
and an approximate resonance width given by
\beq{lrw}
\Delta \Lambda=4 \sqrt{\frac{\beta_1 \beta_2}{3 C \omega}}.\eeq

Another clear aspect of the dynamics in the first-order resonance cases refers to the free eccentricities. The back transformations to original variables, say $\bv q$, are given by series 
of the form $\sum_k {\bv q}^{(k)}$ in terms of the normalising coordinates given by
\beqa{solss}
{\bv q}^{(0)} &=& {\bv q}', \\
{\bv q}^{(1)} &=& \{{\bv q}',\chi_1\}, \\
{\bv q}^{(2)} &=& \{{\bv q}',\chi_2\} + \frac12 \{\{{\bv q}',\chi_1\},\chi_1\}, \\
\vdots &=& \vdots
\eeqa
Using the generating functions obtained above we get in terms of Poincar\'e variables
\beqa{emP1t}
x_1&=& x'_1 - \frac{\partial \chi_1}{\partial y'_1} = x'_1 + \frac{\alpha}{\omega}, \label{xFFP1} \\
y_1&=& y'_1 + \frac{\partial \chi_1}{\partial x'_1} =  y'_1, \\
x_2&=& x'_2 - \frac{\partial \chi_1}{\partial y'_2} = x'_2 + \frac1{\omega} \left( \beta_1 + \beta_2 \cos \lambda' \right), \\
y_2&=& y'_2 + \frac{\partial \chi_1}{\partial x'_2} =  y'_2 +  \frac{\beta_2}{\omega} \sin \lambda' , \\
x_3&=& x'_3 - \frac{\partial \chi_1}{\partial y'_3} = x'_3 + \frac{\gamma}{\omega} \cos \lambda', \\
y_3&=& y'_3 + \frac{\partial \chi_1}{\partial x'_3} =  y'_3 + \frac{\gamma}{\omega} \sin \lambda' , \label{xFFP3}
\eeqa
where the `new' $(x'_i,y'_i)$ represent the free eccentricity oscillations 
$$
x'_i = c_i \cos \omega (t-t_0), \quad y'_i = d_i \sin \omega (t-t_0).$$
The amplitudes $(c_i,d_i)$ are fixed by initial conditions. These relations show how the transformation to the Laplace normal form, aimed at removing 
non-resonant terms 
depending on $q_i$, automatically shifts the eccentricity vectors to the forced equilibria. The back transformations (\ref{xFFP1}-\ref{xFFP3}), when imposing the equilibrium conditions $\lambda'=0$ or $\lambda'=\pi$, are therefore in perfect agreement with solutions \eqref{XE1}.

%
%

\subsection{Non-linear normalisation}
Let us now adopt the book-keeping criterion introduced to treat the general cases (\ref{HBunoS}-\ref{HBdueS}) in which also large eccentricities are admitted and so consider the starting Hamiltonian \eqref{resoH2}
$$
\H = \H_0 + \e \H_1
$$ 
where now the whole integrable part (including quadratic terms in the actions) is used as unperturbed Hamiltonian with respect to which to normalise \citep{bat-mor-ApJ-2015}
$$
\H_0 = \omega\Gamma - \left(\frac32  A \Gamma^2  + 3 B \Lambda \Gamma \right)
$$
and
$$
\H_1 = \H_{res} - \frac32 C \Lambda^2.$$
Comparing with \eqref{resoH1}, this is equivalent to not considering anymore $\omega$ `fast'. 

The solution of the first homological equation \eqref{eq:HomologicGen1} now provides the generating function
\beqa{newchi}
\chi_1 ({\bv Q}',{\bv q}') &=& 
-\frac{\alpha \sqrt{2 Q'_1} \sin{q'_1} + \beta_1 \sqrt{2 Q'_2} \sin{q'_2}}
{\omega - 3 A \Gamma' - 3 B \Lambda'} + \nn
&& -\frac{\beta_2 \sqrt{2 Q'_2} \sin{(q'_2-\lambda')} + \gamma \sqrt{2 Q'_3} \sin{(q'_3-\lambda')}}
{\omega - 3 (A-B) \Gamma' - 3 (B-C) \Lambda'}
\eeqa
and we get 
$$\K_1=- \frac32 C \Lambda^2$$ 
implying the successful elimination of the harmonics of order $\e$. In this case however, in view of the non-linear nature of $\H_0$, this implies the appearance in $\mathcal{H}^{(2)}({\bv Q}',{\bv q}')$  of harmonics of order $\e^2$ with several new angle combinations \citep{MoGi97,MO02}. In particular, in $\K_2$ they appear terms with $q_2-\lambda$ and $q_3-\lambda$ that can be used to describe `beats' between $\omega$ and $\omega_L$. For the sake of the present work its detailed form is not relevant and we limit ourself only to report the change of the librating term in $\cos \lambda$ with respect to \eqref{rnfl}
$$
\frac{\beta_1 \beta_2}2 \left(\frac1{\omega - 3 A \Gamma' - 3 B \Lambda'}+ 
\frac1{\omega - 3 (A-B) \Gamma' - 3 (B-C) \Lambda'} \right) \cos \lambda'
$$
providing a first approximation of the effect of the non-linear terms on the pendulum dynamics. 
In particular, we can see how is removed the divergence in \eqref{fola} 
and \eqref{lrw} when $\omega \to 0$.
On the other hand, it is very 
interesting to compare the back transformation of the coordinates given by the generating function \eqref{newchi}.  
In analogy with (\ref{xFFP1}-\ref{xFFP3}) we get now
\beqa{emP2t}
x_1= x'_1 &+& \frac{\alpha}{\omega - 3 A \Gamma' - 3 B \Lambda'}  \nn
&+& \frac{3A(\alpha y_1' + \beta_2 y_2') y_1'}{(\omega - 3 A \Gamma' - 3 B \Lambda')^2} 
+ \frac{3(A-B)\left[(\beta_2 y_2' +  \gamma y_3') \cos \lambda' - (\beta_2 x_2' +  \gamma x_3') \sin \lambda' \right]y_1'}{(\omega + 3 (B-A) \Gamma' - 3 (C-B) \Lambda')^2}, \nn
x_2= x'_2 &+& \frac{\beta_1}{\omega - 3 A \Gamma' - 3 B \Lambda'} +
\frac{\beta_2 \cos \lambda'}{\omega + 3 (B-A) \Gamma' - 3 (C-B) \Lambda'} \nn
&+& \frac{3A(\alpha y_1' + \beta_2 y_2') y_2'}{(\omega - 3 A \Gamma' - 3 B \Lambda')^2} 
+ \frac{3(A-B)\left[(\beta_2 y_2' +  \gamma y_3') \cos \lambda' - (\beta_2 x_2' +  \gamma x_3') \sin \lambda' \right]y_2'}{(\omega + 3 (B-A) \Gamma' - 3 (C-B) \Lambda')^2}, \nn
x_3= x'_3 &+& \frac{\gamma \cos \lambda'}{\omega + 3 (B-A) \Gamma' - 3 (C-B) \Lambda'}  \nn
&+& \frac{3A(\alpha y_1' + \beta_2 y_2') y_3'}{(\omega - 3 A \Gamma' - 3 B \Lambda')^2} 
+ \frac{3(A-B)\left[(\beta_2 y_2' +  \gamma y_3') \cos \lambda' - (\beta_2 x_2' +  \gamma x_3') \sin \lambda' \right]y_3'}{(\omega + 3 (B-A) \Gamma' - 3 (C-B) \Lambda')^2}, \nonumber
\eeqa
for the $x_i$ and similar expressions for the $y_i$. Imposing again the equilibrium values $x_i'=0, y_i'=0$ (which imply \eqref{LEGE}) and $\lambda'=0,\pi$, we can check that we re-obtain system \eqref{YEQ2}. Therefore, also in this case the procedure of Lie-transform normalisation produces the correct shift to the forced equilibria.

\section{Applications}\label{AppSec}

There are two case-studies that happily perfectly fit the two main occurrences of relative equilibria 
illustrated above in Section \ref{sec:eql}: the Galilean system is the standard example of Laplace resonance 
close (even if not exactly) at a de Sitter-Sinclair equilibrium; the exo-planetary system Gliese 876 is instead 
well described by one of the new-de Sitter equilibria. In the light of the theory discussed above, we illustrate these two examples in the following subsections. There is no claim to rigor: the aim is just to show the main aspects of these results. 

%
\subsection{The Galilean system}

We use the mass-parameters of the Galilean system to investigate the corresponding family of equilibria. We remark that the model is highly idealised: in particular, the oblateness of Jupiter is neglected. 

\begin{figure}[h]
\centering
\hglue-1cm
\includegraphics[width=12.7truecm]{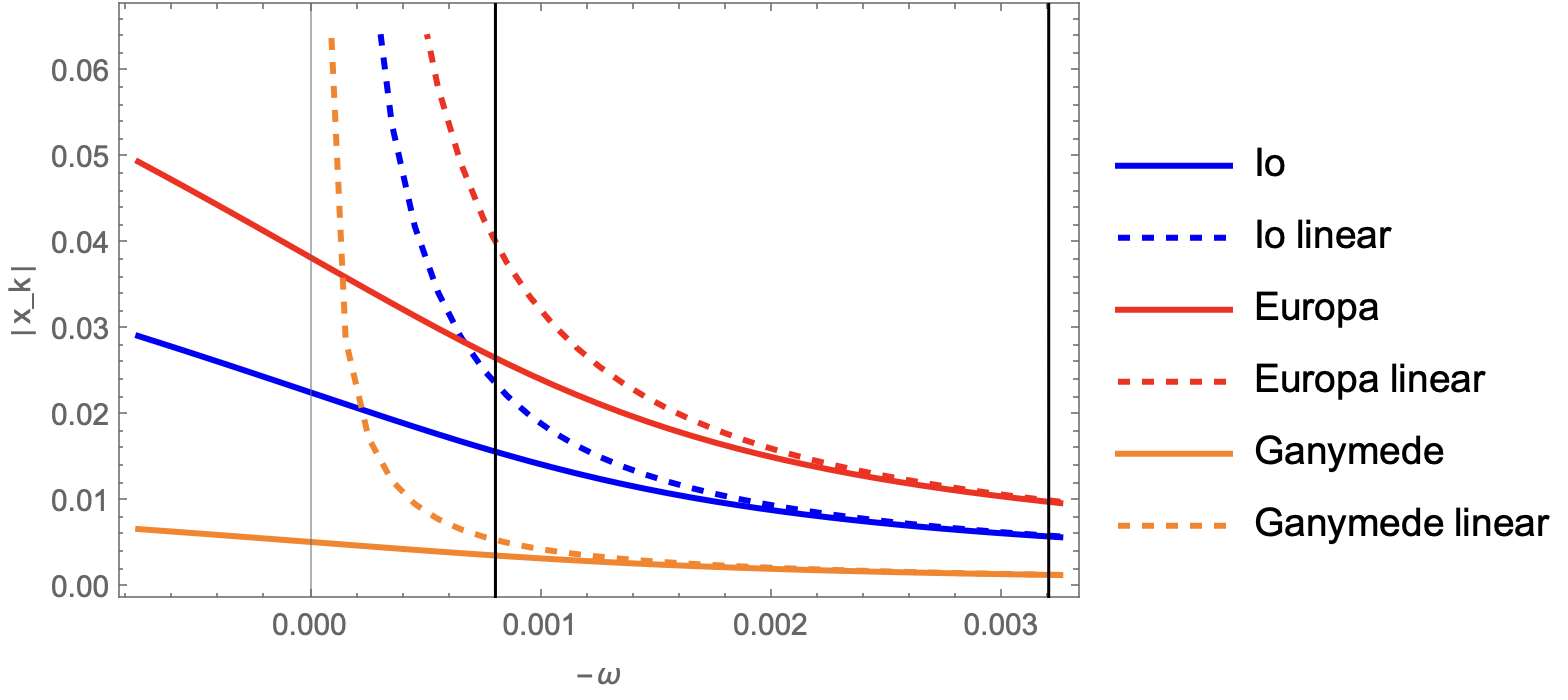}
\caption{De Sitter-Sinclair equilibria for the Galilean system: continuous lines correspond to solution \eqref{XE1E}; dashed lines to the approximate solutions  (\ref{YE1}-\ref{XE1}). The verse of the proximity parameter $\omega$ is reversed. The vertical thick lines give the actual observed value ($\omega = -0.00325$) and the instability threshold ($\omega = -0.00078$) predicted by applying \eqref{LUNST}.}
\label{fig:Gali1}
\end{figure}

\begin{figure}[h]
\centering
\hglue-1cm
\includegraphics[width=8.7truecm]{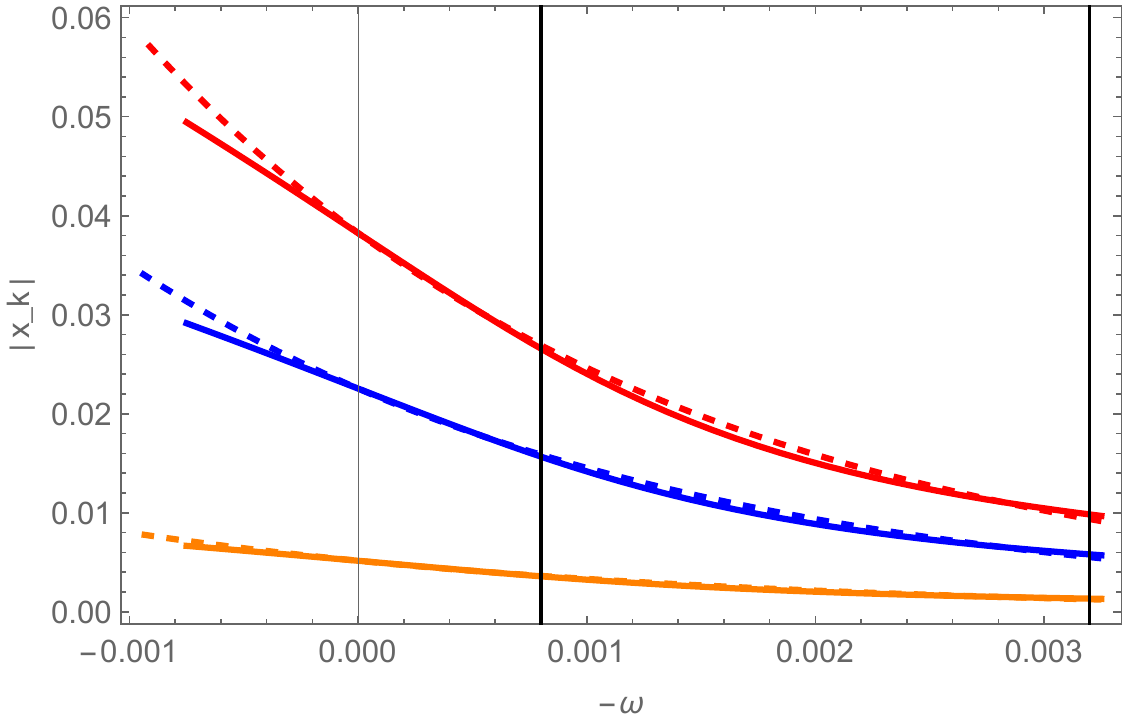}
\caption{De Sitter-Sinclair equilibria for the Galilean system: comparison between the analytical solutions (\ref{XE1E}, continuous lines)  and the exponential fit (\ref{XEXP1}, dashed lines). The color code is the same as in Fig.\ref{fig:Gali1}}
\label{fig:Gali3}
\end{figure}

In Fig.\ref{fig:Gali1} we plot both the exact  and approximate solutions (respectively \eqref{XE1E} and (\ref{YE1}-\ref{XE1})) to show their very different behaviour when $\omega\to 0$. 
The verse of the proximity parameter is reversed in order to easily compare these solutions with the classical results published in \cite{YoPe} and \cite{HeCons}. We refer to eq.\eqref{omefro} to recall the connection between the proximity parameter and the resonance offsets used in those references: 
in particular, \citet{HeCons} adopts the same units of measures used here \citep{GP-CMDA-2021}, showing a good agreement with our approximate solutions.  In Fig.\ref{fig:Gali3} we also perform the comparison between the analytical solutions \eqref{XE1E} and the exponential fit \eqref{XEXP1}. 

\begin{figure}[h]
\centering
\hglue-1cm
\includegraphics[width=9.2truecm]{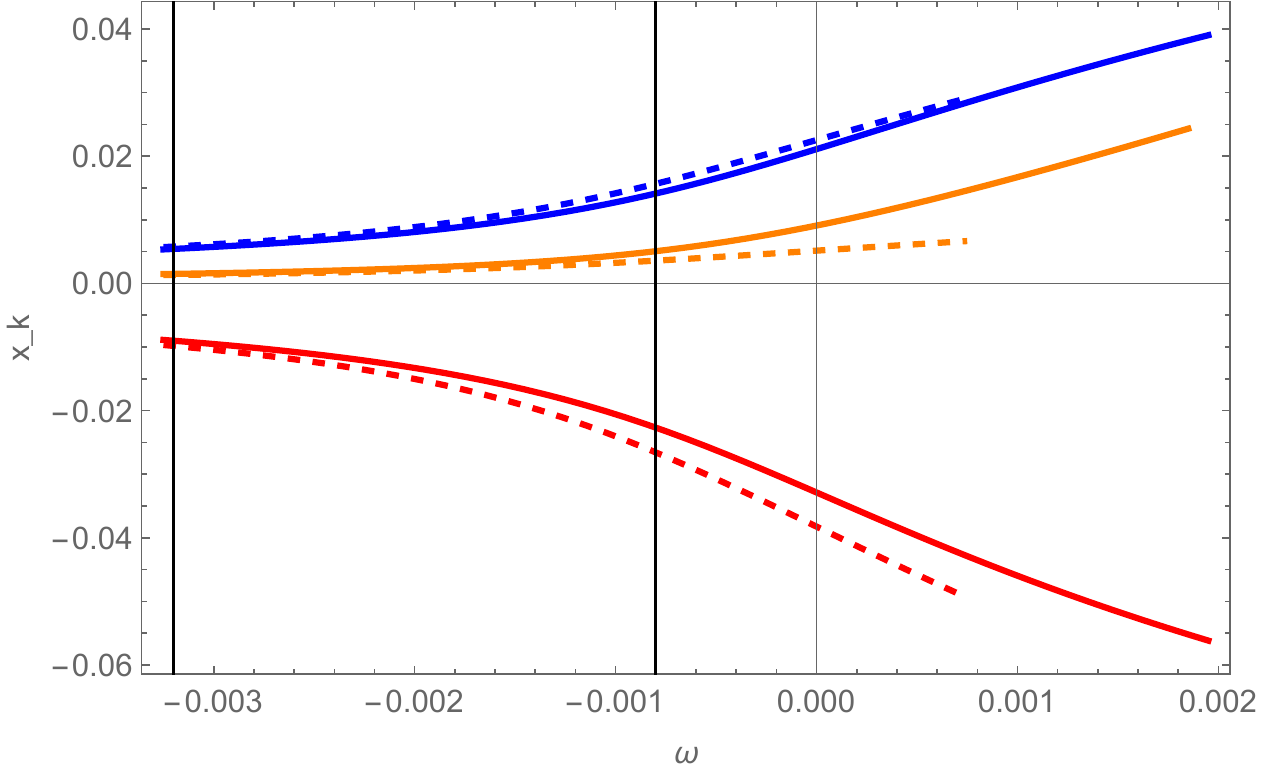}
\caption{De Sitter-Sinclair equilibria for the Galilean system: comparison between the analytical solutions (\ref{XE1E}), dashed lines, and exact solutions obtained by a root-finding method of the complete model including the 2nd-order terms in eccentricity (continuous lines). The color code is the same as in Fig.\ref{fig:Gali1}.}
\label{fig:Gali4}
\end{figure}

In Fig.\ref{fig:Gali4} we compare the analytical solutions \eqref{XE1E} with exact solutions obtained by a root-finding method of the complete model including 2nd-order terms in eccentricity. The verse of the proximity parameter is now the standard one; moreover, we can also see the sign of each solution which, according to the definition of Poincar\'e variables, determines the value of the libration centre for each resonant angle. 

Finally, in the following tables we report the eigenvalues of the stability matrix at equilibrium: in Table \ref{Tp} we have the case very close to the actual state of the Galilean system ($Q_5 = 1.2259, Q_6 = 4.3199, \omega = -0.00325$), respectively using the first and the 2nd-order solutions;  in Table \ref{T3}, that corresponding to a configuration deep in resonance, ($Q_5 = 1.2257, Q_6 = 4.3127, \omega = -0.00075$), which according to the first-order theory should be (and indeed is not) dynamically unstable. 

\begin{table}
\begin{tabular}{@{}lccccccc@{}}
  \hline
eigenvalues   &  Analytical	 1st-order	       &  Analytical 2nd-order	 &  Numerical  &  \\
 \hline
 $\sigma_2$   &  $0.00111 i $     & $0.00096 i $ & $0.00086 i $&\\
 \hline
$\sigma_4$    &  $0.00284 i $     & $0.00392 i $ & $0.00390 i $& \\
\hline
$\sigma_6$    &  $0.00352 i $     & $0.00423 i $ & $0.00400 i $&\\
\hline 
$\sigma_8$    &  $0.00369 i $    &  $0.00534 i $ & $0.00463 i $ & \\
\hline\hline
\end{tabular}
\caption{Galilean system (observed): $Q_6 = 4.3199, \omega = -0.00325$.} \label{Tp}
\end{table}

\begin{table}
\begin{tabular}{@{}lccccccc@{}}
  \hline
eigenvalues   &  Analytical		 1st-order		         	         &  Analytical 2nd-order &   Numerical  &\\
 \hline
$\sigma_2$   &  $0.00448 $     & $0.00061 i $   & $0.00070 i $& \\
 \hline
$\sigma_4$    &  $0.00056 i $  & $0.00082 i $   & $0.00129 i $& \\
\hline
$\sigma_6$    &  $0.00075 i $   & $0.00141 i $  & $0.00161 i $& \\
\hline 
$\sigma_8$    &  $0.00353 i $  & $0.00333 i $  &  $0.00213 i $ & \\
\hline\hline
\end{tabular}
\caption{Galilean system (closer to resonance): $Q_6 = 4.3127, \omega = -0.00075$.} \label{T3}
\end{table}

\subsection{GJ-876}
The system Gliese-Jahrei{\ss} 876 (GJ-876) is very important because it was the first case of discovery of mean motion-resonance outside our Solar System 
\citep{marcyeal} and of multi-resonance among three planets \citep{RG876}. 
In Table \ref{T2} we list the nominal elements reported in \cite{BEN}. Semi-major axes and eccentricities are quite well known, with $e_1=0.2531$ for the first planet in the chain, planet-c, a Jupiter-like with a period of 30 days. A second Jupiter, planet-b (since it was the first to be discovered) has a period of 61 days and finally there is a Uranus-like object, planet-e, with a period of 124.5 days. 

\begin{table}

    \begin{tabular}{@{}llllcccc@{}}
  \hline\hline
semi-major axis [in unit of $a_1$]                                               		&\rm{Planet c}                   & \rm{Planet b}             	& \rm{Planet e} & & & & \\
\hline

 nominal							                                 		&  1                  			& 1.6074               		&       2.5840   & & & & \\
 de Sitter-Sinclair	                            						&  1                     		& 1.6093                  		&   	 2.6350  & & & & \\
 new de Sitter 		                            						&  1                     		& 1.6052                  		&   	 2.5829  & & & & \\
 \hline\hline
 eccentricity/resonant angles 						& $e_1$		& $e_2$		& $e_3$ 	& $q_1$ \quad & \quad $q_2$& $q_3$ & $q_4$ \\
 \hline
 
observed	                         									&  0.2531 				& 0.0368 				& 	 0.0310  & 0 \quad & \quad $0$ & rotating? & $0$ \\
de Sitter-Sinclair  ($X^A_k(\pi)$)                                  			&  0.0480				& 0.0047				&       0.0064  & $\pi$ \quad & \quad $0$ & $0$ & $\pi$ \\
de Sitter-Sinclair  ($X^A_k(0)$)			                                		&  0.0524				& 0.0104				&       0.0023  & $\pi$ \quad & \quad $0$ & $0$ & $0$ \\
new de Sitter ($X_k(\pi)$)                                  				&  0.2657				& 0.0737				&       0.0117  & 0 \quad & \quad $\pi$ & $\pi$ & $\pi$ \\
new de Sitter ($X_k(0)$)			                                		&  0.2546				& 0.0366				&       0.0381  & 0 \quad & \quad $0$ & $\pi$ & $0$ \\
\hline\hline
\end{tabular}
\caption{Mean nominal orbital elements of the 3 main planets in GJ-876 and resonant angles according to 
\cite{BEN} compared with predictions from the model.} \label{T2}
\end{table}

\begin{figure}
  \includegraphics[scale=0.3]{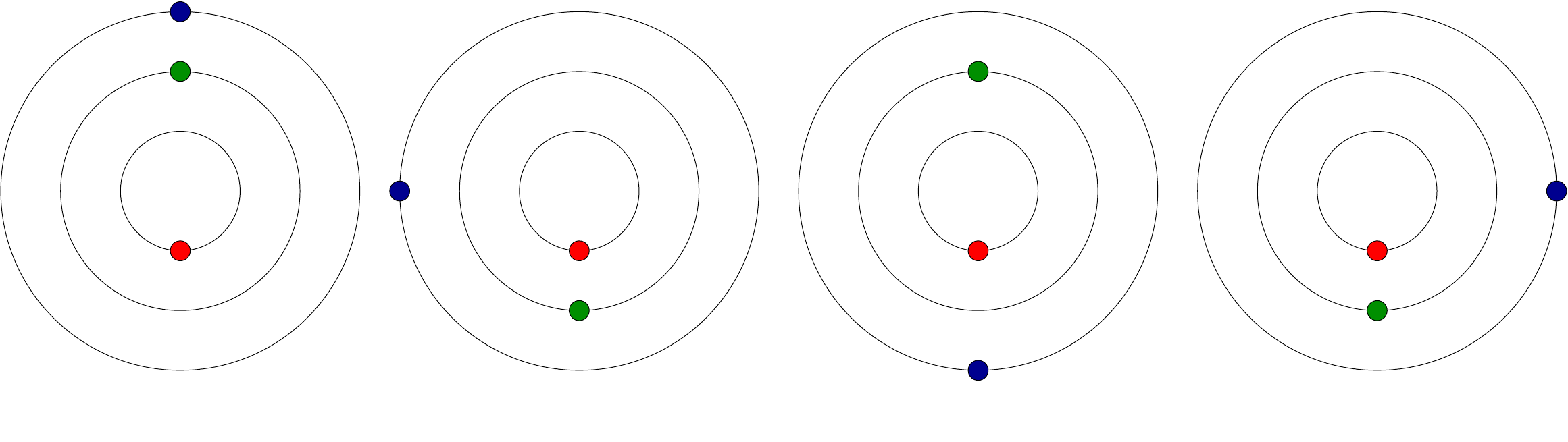}\\
  \includegraphics[scale=0.3]{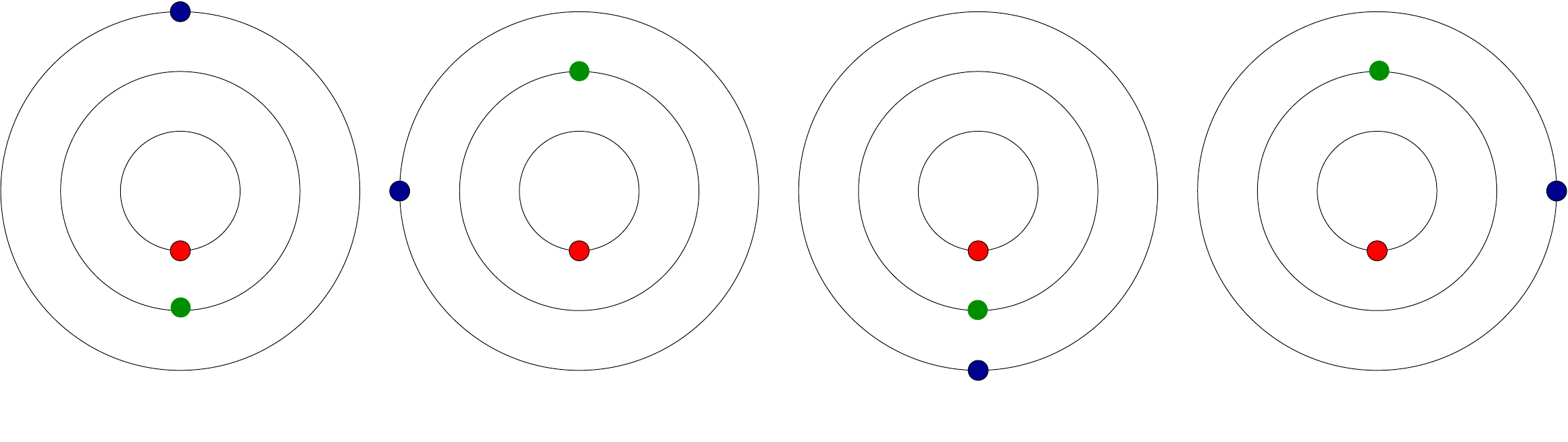}
  \caption{Idealised pictures of the Galilean system (top panel) compared with GJ-876 (lower panel). Each snapshot corresponds to 
  a revolution period of Io or planet-c (red dots); in the upper plots Europa (green dots) and Ganymede (blue dots) can 
  be in conjunction with Io one at a time; in the lower plots, planet-b (green dots) and planet-e (blue dots) can together be in conjunction with planet-c.}
  \label{fig:RG}
\end{figure}

A remarkable aspect of this configuration concerns the possibility to have \sl triple conjunctions: \rm in the lower panel of Fig. \ref{fig:RG} 
it can be seen that these configurations are allowed \citep{RG876}
with planet-c and -b at periastron ($\lambda_1=\lambda_2=p_1=p_2=0$) and planet-e at apastron ($\lambda_3=0,p_3=\pi$). 
On the other hand, in the Galilean case Europa and Ganymede can 
only  be in conjunction with Io one at a time.

GJ-876 is then the prototype of systems in which the new de Sitter equilibria appear to be in very good agreement with observation. It is clearly stimulating to try to understand these results in the 
framework of the theory of resonant capture \citep{CMBR18,PiBaMo} and also to get clues about the structure of the other systems for which less accurate 
informations are available. 
 
\begin{figure}[h]
\centering
\hglue-1cm
\includegraphics[width=9.5truecm]{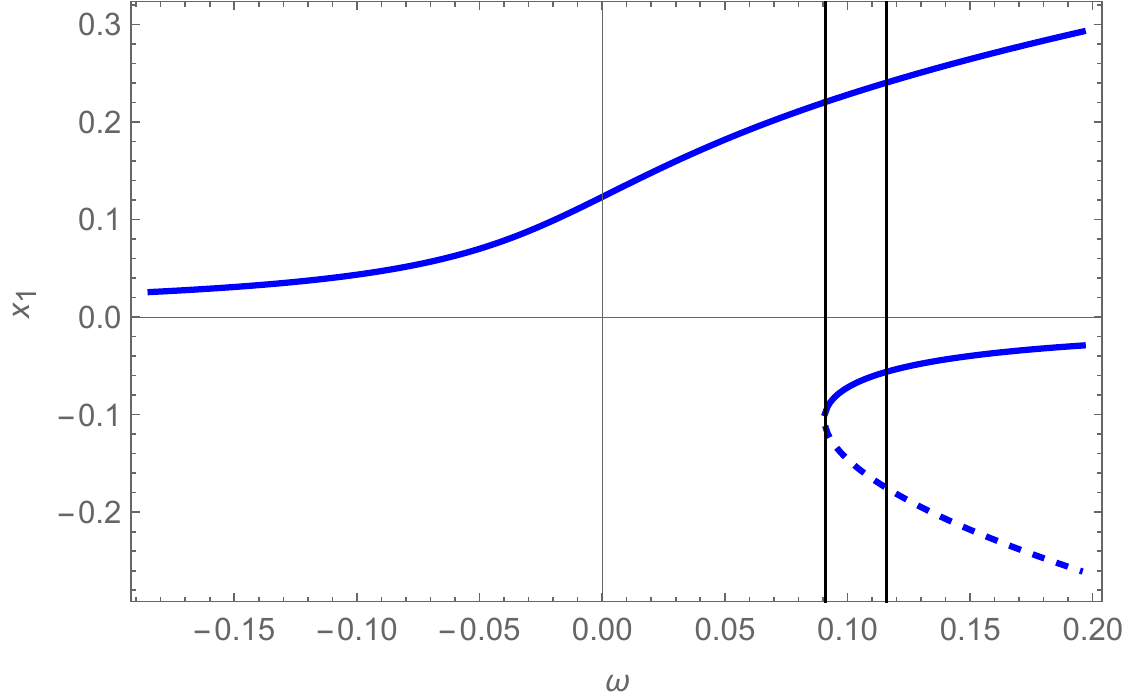}
\caption{New deSitter for the GJ-876 system: $X_1$ solutions with $\lambda = 0$. The vertical thick lines give the actual observed value ($\omega = 0.1159$) and the bifurcation threshold ($\omega = 0.0909$).}
\label{fig:G1}
\end{figure}

\begin{figure}[h]
\centering
\hglue-1cm
\includegraphics[width=9.5truecm]{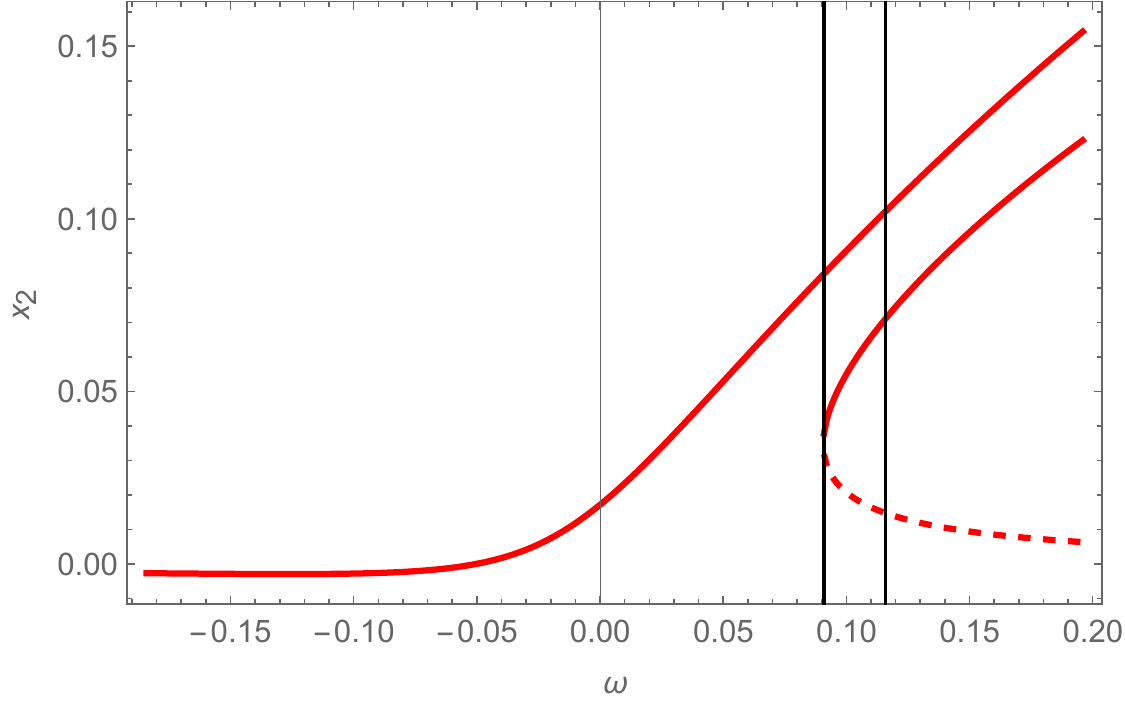}
\caption{Same as Fig.\ref{fig:G1}: $X_2$ solutions.}
\label{fig:G2}
\end{figure}

\begin{figure}[h]
\centering
\hglue-1cm
\includegraphics[width=9.7truecm]{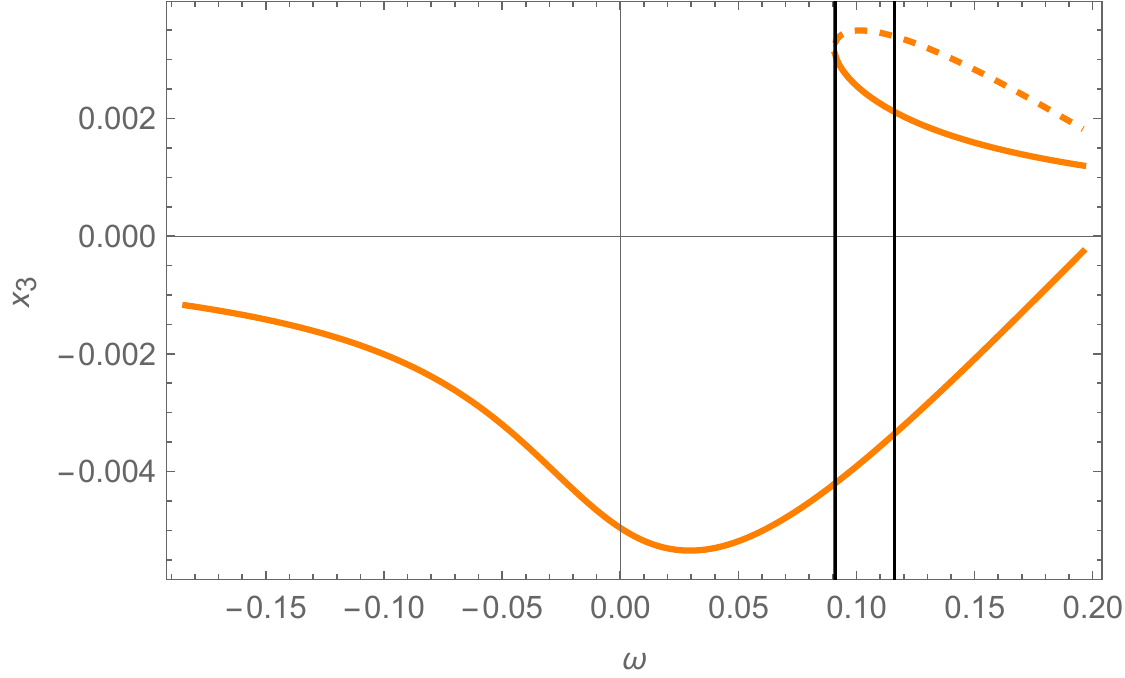}
\caption{Same as Fig.\ref{fig:G1}: $X_3$ solutions.}
\label{fig:G3}
\end{figure}

The bifurcation value of the proximity parameter, when upgraded with the complete 2nd-order terms is
$$ \omega_b' = \omega_b + 2 \alpha_2 = 0.0909. $$
Inserting the observed data, it appears that the current status of the system corresponds to  $\omega = 0.1159$, 
so that it is beyond the bifurcation threshold. Therefore, the solution which best describes the actual system is 
indeed the stable branch after the saddle-node bifurcation of the sequence with libration center of the Laplace argument at $\lambda = q_4 = 0$. In Figs. \ref{fig:G1}-\ref{fig:G3} we plot the exact 2nd-order solutions computed with a root-finding method. We can see a close similarity with the equilibria obtained by \cite{B15} in the case of the 2-planet MMR case. The bifurcation threshold computed above seems to be quite accurate: note that also here the usual verse of the abscissa-axis is used. Comparing with values reported in Table \ref{T2}, we can verify how also the apsidal configuration is correctly reconstructed by the new de Sitter family with $\lambda = 0$.

\section{Conclusions}\label{CSec}

We can summarise the results of the present work with the following remarks:

\noindent
1. The sequence of Laplace-resonant equilibrium configurations can be parametrised 
by a single frequency (the `resonance proximity parameter') 
depending on the two integrals of motion admitted by the Hamiltonian model. 

\noindent
2. The equilibrium is dynamically (Lyapunov) stable all along the sequence up to a threshold 
determining a saddle/node bifurcation with two additional stable/unstable sequences. 

\noindent
3. The range of models in which the resonance proximity parameter is close to vanishing 
require a proper ordering of the Hamiltonian series which impacts on the construction of the 
resonant normal form governing the dynamics around the equilibrium.

 The simple model in which only first-order resonant terms are considered allows us to get 
 explicit solutions for the equilibrium sequence and the bifurcation threshold. It is 
 qualitatively correct also for what concerns the apsidal configurations. To get more accurate 
 quantitative results, the inclusion of higher-order terms requires numerical 
 root-finding methods. 
 
 There are at least two straightforward extensions of this work that can be proposed for further study: 
 first, a systematic investigation of the equilibrium sequences by varying the physical parameters (in particular the mass-ratios); second, to exploit the non-linear normal form to study higher-order harmonics in the perturbations. 
 
\section*{Acknowledgements}
I acknowledge the  
ASI Contract n. 2023-6-HH.0 (Scientific Activities for JUICE, E phase)
and partial support from GNFM/INdAM and INFN (Sezione di RomaII).
\section*{Compliance with ethical standards}

{Conflict of interest:} The author declares that he has no conflict of interest.

\bibliographystyle{spbasic}
\bibliography{juice1} 
\end{document}